\documentclass[default,iicol]{sn-jnl}

\jyear{2021}%

\theoremstyle{thmstyleone}%
\theoremstyle{thmstyletwo}%
\usepackage{flushend}
\theoremstyle{thmstylethree}%
\usepackage{natbib}
\setcitestyle{authoryear,round}

\begin{document}

\title[]{A new model of quasar mass evolution.}


\author*[1,2,3,5]{\fnm{Zheng Li} }\email{lizheng@xao.ac.cn}

\author[1,2,3,5]{\fnm{Ming Zhang}}

\author[4]{\fnm{Qiu-He Peng}} 
\author[1,2,3,5]{Xiang Liu}
\affil*[1]{\orgdiv{Xinjiang Astronomical observatory}, \orgname{Chinese Academy of Sciences}, \orgaddress{\street{150 Science 1-Street}, \city{Urumqi}, \postcode{830011}, \country{China}}}

\affil[2]{\orgdiv{Key Laboratory for Radio Astronomy}, \orgname{Chinese Academy of Sciences}, \orgaddress{\street{2 West Beijing Road}, \city{Nanjing}, \postcode{210008}, \country{China}}}

\affil[3]{\orgname{University of Chinese Academy of Sciences}, \orgaddress{\street{19A Yuquan Road}, \city{Beijing}, \postcode{100049}, \country{China}}}


\affil[4]{\orgdiv{ School of Astronomy and Space Science}, \orgname{Nanjing University}, \city{Nanjing}, \postcode{210023}, \country{China}}

\affil[5]{\orgdiv{Key Laboratory of Radio Astrophysics in Xinjiang Province}, \city{Urumqi}, \postcode{830011}, \country{China}}

\abstract{Magnetic monopoles have been a trending topic among
physicists and astronomers since the 1930s. Researchers have been
working hard to find evidence of magnetic monopoles in
laboratories. The existence of magnetic monopoles can rationally
explain the stability of charges, the quantization of charges, the
structure of leptons, the unified composition of leptons and
hadrons, and the symmetry of leptons and quarks. The presence of
these mysterious particles in the universe could have significant
implications for theoretical physics and astrophysics. The Grand
Unified Theory has also predicted the existence of magnetic
monopoles, which is interestingly implied by some astronomical
observations. Noticing that the growth of supermassive black holes
in the early universe is an increasingly challenging difficulty
faced by astronomers, here we argue that it could be solved with
the help of magnetic monopoles. As suggested by Peng et al. in
1986, quasars containing magnetic monopoles at the center can
continuously catalyze the decay of protons to release energy. We
examine this model by using quasar data from the Sloan digital
sky survey. It is shown that the initial mass distribution of
quasars derived from the magnetic monopole model exhibits a
Gaussian distribution. At the same time, the initial mass function
is also slightly higher than previously expected, which
could be verified by future observations.}




\maketitle

\section{Introduction}

Since Maarten Schmidt discovered the first quasar in 1963
\cite{1963Natur.197.1040S}, much progress has been made in this
field. Thousands of research articles about quasars have been published
each year in the past decade
 \citep{1983ApJ...269..352S,1995AJ....109.1498H,2001ApJ...551L..23P,2008PhDT........13S,2018PASJ...70S..37G,2019MNRAS.484L..97S}.
It is now commonly held that quasars live in galaxy
centers
 \citep{1995ARA&A..33..581K,2001AIPC..586..363K,2004ApJ...601..676V,2018MNRAS.481.3118M,2019MNRAS.484.2851L}.
The observations of nearby galaxies have revealed a tight
correlation between the mass of quasars and various parameters of
their host galaxies, such as the stellar velocity dispersion or
the mass of the bulge \citep{2011Natur.480..215M,2011MNRAS.412.2211G,2013ARA&A..51..511K,2013ApJ...764..184M}.
 These relations strongly imply a co-evolution among quasars
and host galaxies. This means that quasars play an essential role in
galaxy formation and evolution, and it is thereforce crucial to measure
the masses of quasars accurately. Assuming a virialized equilibrium for
the broad line region (BLR) featured by broad $H_\alpha$, $H_\beta$, $M_{g
\uppercase\expandafter{\romannumeral2}}$ and $C_{\uppercase\expandafter{\romannumeral4}}$ emission lines, one can
derive the masses of quasars \citep{2011ApJS..194...45S,2015ApJ...806..109J}.

At present, more than 200 high-redshift ($z>6$) quasars have been
discovered
\citep{2006AJ....132.2127J,2009ApJ...702..833K,2015ApJ...806..109J,2016ApJ...828...26M,2017ApJ...849...91M,2018Natur.553...51M,2018Natur.553..473B,2019arXiv190304078F}.
The masses of these quasars are almost universally larger than
$10^{9}M_\odot$. Some of them even reach
$10^{10}M_\odot$\citep{2015Natur.518..512W}. How these quasars
grew so quickly over a short period (approximately 700 Myr after the Big Bang)
is still unknown. Much work has been devoted to solving this
problem. For example, the Pop I and Pop II seed scenarios assume
that a $\sim 100 M_\odot$ black hole could be formed as a seed
after the collapse of a single massive star \citep{2002Sci...295...93A, 2002ApJ...564...23B} at redshift
10 --- 20. It then grows through sub-Eddington accretion \citep{2018ApJ...855..138K}. In this case, the growth time of the
black hole is only 0.76 Gyr, which is determined by its redshift
of $z \sim 6$ --- 20. However, astronomers recently observed a
supermassive quasar that was larger than $10^{10}M_\odot$. It
would need an accretion time of $> 10^{9} yr$ even through
Eddington accretion, which conflicts with its age. Thus the accretion models
are confronted with a serious problem. Some researchers have argued that the accretion
could be super-Eddington, especially for geometrically thick
accretion disks  \citep{2015MNRAS.451.1964S}. However, a super-Eddington accretion would exert a centrifugal force on the
rotating ring of the accretion flow, which means the
super-Eddington stage could not be maintained for a long
time\citep{1998bhad.conf.....K}. \par

An alternative model, the ``direct collapse black hole'' model
(DCBH), assumes that the SMBH seed is of mass $\sim10^{5}M_\odot$
and is directly formed at the redshift $z \sim 10$ --- 20 through a
rapid collapse. This model is somewhat supported by the
observations of high luminosity Ly$\alpha$ emitting galaxies at
high redshifts in recent years
\citep{2013ASSL..396..293H,2018MNRAS.476..366B}. Note that the
DCBH model needs some extreme conditions such as a metal-free gas
or a high-temperature environment ($\sim10^{4}K$). Unfortunately,
some simulations indicate that the collapsing $H_2$ clouds cool
rapidly to temperatures lower than  300 K \citep{2010MNRAS.402.1249S}. Additionally,
the latest simulations further show that black holes in
the early universe were not immersed in the lowest point of
gravitational potential, which is contrary to previous simulation
results. In fact, they could be kicked away from the lowest point
of the gravitational potential, slowing down the accretion rate.
Thus, this does not allow for a rapid increase in the accreted mass,
which poses more serious problems for current black hole growth
theories \cite{2021MNRAS.508.1973M}.\par

In this study, we propose an alternative mechanism to explain the growth of
supermassive black holes, which is different from the current mainstream models. The
structure of our paper is organized as follows. In Section \ref{sec1}, we briefly
introduce the magnetic monopole model. The difference between this model and the
previous mainstream accretion disk model is described. In Section \ref{sec2}, we
describe the selection of our quasar sample, which includes 105,783 quasars with the mass
and redshift available \citep{2011ApJS..194...45S}. Section \ref{sec4} presents
the initial masses of the quasars calculated from our magnetic monopole model.
Finally, Section \ref{sec5} offers our conclusion.

\section{A brief description of the magnetic monopole model}\label{sec1}

In the 1970s, a new concept called the Rubakov Callen effect
(hereafter referred to as the RC effect) was proposed
\citep{1981ZhPmR..33..658R,RUBAKOV1982311,RUBAKOV1983240,PhysRevD.25.2141},
which stated that magnetic monopoles can catalyze nucleon decay
into leptons. In 1982, Guth proposed that a tiny amount of
magnetic monopoles may be produced by the violent oscillation and
thermal fluctuation of the Higgs field during the phase transition
of the primordial universe, which is super-hot ($k T  > 10^{15}$
GeV) \citep{1982botu.conf...25G}. In 1985, Peng et al. suggested
that the magnetic monopoles deposited at the center of active
galactic nuclei (AGN) can continuously catalyze the decay of
protons into leptons to release energy so that the RC effect can
act as the major energy source for AGN. These magnetic monopoles
would prevent the center regions of AGN from collapsing to infinity
to form black holes. Their theory points to the existence of new
compact objects different from black holes
\citep{1986ASSL..121..663P,1998ASPC..151..119P,
2001ApJ...551L..23P}. In this framework, the gravitational effects
of surrounding areas of AGN containing magnetic monopoles is
similar to that of black holes. However, the supermassive
celestial bodies containing magnetic monopoles have neither
black hole horizons nor singularity. Because magnetic monopoles
can catalyze the decay of nucleons and the reaction rate is
proportional to the square of the density of matter in the core
region, the energy power (called the RC luminosity) will be large
enough to prevent a celestial body from gravitationally
collapsing thanks to the intense radiation pressure.\par

Due to the inefficiency of the Hawking radiation, the mass
of a quasar will continuously increase in the framework of the
standard accretion disk model. However, in the magnetic monopole
model, the mass of the supermassive quasar can decrease because
catalytic proton decay reactions are triggered by magnetic
monopoles at the center. The detailed pattern of the mass
evolution will be derived and compared with that of the
accretion disk model below.

\begin{figure}[h]
 \centering
     \includegraphics[scale=0.5]{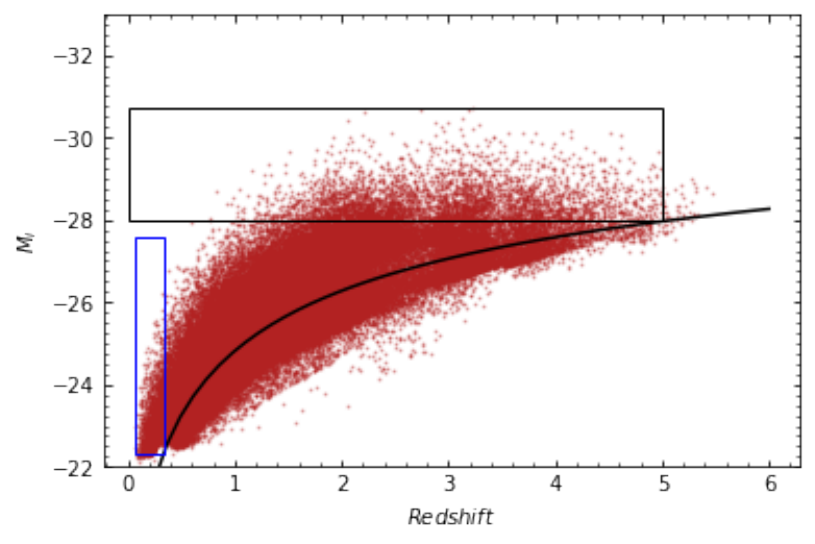}
     \caption{$i$-band absolute magnitude versus the redshift for all the observed
             quasars. The black line is the flux limit after K-correction.
             Our final samples are limited in the black rectangle region, which
             corresponds to $0<z<5$ and $M_{i}<-27.5$. The blue rectangle
             shows 222 objects with a mass larger than $10^{9}M_\odot$.
             The data are taken from \cite{2011ApJS..194...45S}.}

     \label{image_1}
\end{figure}

\section{Sample selection}\label{sec2}

The SDSS DR7 data sample we used contain 105783 quasars with
redshifts in a range of $z \sim 0-5$. Some useful parameters, such
as the quasar mass ($M$), the 5100$\AA$ monochromatic luminosity
($L_{5100}$), and the radiation luminosity ($L_{bol}$), are
available. Combining these data with the accretion disk
theory allows us to explore the evolution of quasar mass.
Generally, there are two kinds of models for AGN accretion.
One is a geometrically thin optically thick disk called a thin
disk or bright model \citep{2008MNRAS.391..481S}. The
other is a geometrically thick optically thin disk called a
thick disk model. The difference between them is that the
radiation efficiency of the thick disk is much lower than 0.1, and
a relativistic jet can be launched. The thin disk model can
produce optical and X-ray emission lines, and the radiation
efficiency is $\eta \simeq$ 0.1 --- 1. They are usually bright
quasars with an apparent magnitude of $i < 20.2$, which is
frequently used as a useful criterion for identifying quasars in
sky surveys
\citep{2009ApJS..182..543A,2010AJ....139.2360S,2011ApJS..194...45S}.
They all have emission lines in optic bands, so we use the thin
disk accretion model for them in our study.

Using the data of quasar mass and the 5100 $\AA$ monochromatic
luminosity ($L_{5100}$), we can obtain the accretion rate ($\dot{M}$)
from the thin disk theory as \cite{2013peag.book.....N}
\begin{equation}
 \dot{M}\simeq 2.6\left [ \frac{L_{5100,45}}{cosi} \right ]^{3/2}M_{8}^{-1} \quad M_\odot /{\rm yr}, \label{con:equation1}
\end{equation}
where $M_{8} = M / (10^{8} M_\odot), L_{5100,45}=L_{5100}/10^{45}
{\rm erg/s}$, and $i$ is the inclination angle of the disk with
respect to the line of sight. We take $\cos i = 0.8$ and $\eta =
0.1$ as typical parameters for the sample. Since astronomical
observations are generally carried out using various filters,
there could be gaps between different filter bands. As a result,
we can only measure part of the spectrum. The SDSS survey adopts
the $u,g,r,i,z$ five-filter system. In this study, we uniformly
use the $i$-band data. When we conduct statistical research, we
need to correct the observed wavelength to the same rest-frame
range. This process is called the K-correction, which is defined
as
\begin{equation}
    M_i = m_i - 5 (log_{10} D_{L} - 1) - K(z),
\end{equation}
where $M_i$ is the $i$-band absolute magnitude, $m_i$ is the
corresponding apparent magnitude, and $D_{L}$ is the luminosity
distance. To eliminate the observational selection effect, we cut
the sample according to the magnitude limit of the SDSS telescope,
which can maintain the completeness of the sample statistics as
far as possible. K-correction is performed in the process
\citep{2008AJ....136.1799K,2014A&A...566A...1T}. For example, for
a complete sample of high-luminosity quasars, we choose a sample
with $0 < z < 5$ and $M_{i} < -27.96$. In Figure \ref{image_1},
the rectangular box marks the selected high-luminosity quasar
objects.

The Friedmann equation is \cite{1922ZPhy...10..377F}:
\begin{equation}
    \dot{a}^{2}=H_{0}^{2}\left ( \frac{\Omega_{m}}{a}+\frac{\Omega_{r}}{{a}^{2}}+ \Omega_{\wedge}a^{2}+\Omega_{k}\right ), \label{eq3}
\end{equation}
where $a$ is the cosmic scale factor, $\Omega_{m}$ is the matter
density parameter of the universe, $\Omega_{r}$ is the radiation
density, $\Omega_{\wedge}$ is the vacuum energy density,
$\Omega_{k}$ is the cosmic curvature, and $H_{0}$ is the Hubble
constant. In this paper $\Omega_{m}$ = 0.3,  $\Omega_{\wedge}$ =
0.7, $\Omega_{r}$ = 0, $\Omega_{k}$ = 0, and $H_{0}$ = 69.32
km/(Mpc$\cdot$ s) are adopted from the WMAP9 results. Using
Equation \ref{eq3}, one can calculate the age of the universe:
\begin{equation}
    dt=-\frac{dz}{H_{0}(1+z)^{2}(1+\Omega_{m}z)-z(2+z)\Omega_{\wedge}}. \label{eq4}
\end{equation}

Therefore, using Equation \ref{eq4}, one can obtain the
cosmological age corresponding to a particular redshift. In
principle, we can use the derived accretion rate and the
cosmological age to calculate the initial mass of the quasar.
However, note that the evolution of the accretion disk is a
complex process, accompanied by many other processes such as star
formation, supernova feedback, galaxy mergers, and chemical evolution.
These processes can also change the accretion rate of
quasars. Hopkins and his team performed refined calculations
in this area
\citep{2005ApJ...630..705H,2006ApJS..163....1H,2008MNRAS.391..481S}.
Here, we will not discuss details about the refinement of
the accretion disk model. We will mainly investigate the mass
evolution in the framework of the magnetic monopole model in
detail below.

\begin{figure}[h]
\centering
 \includegraphics[scale=0.5]{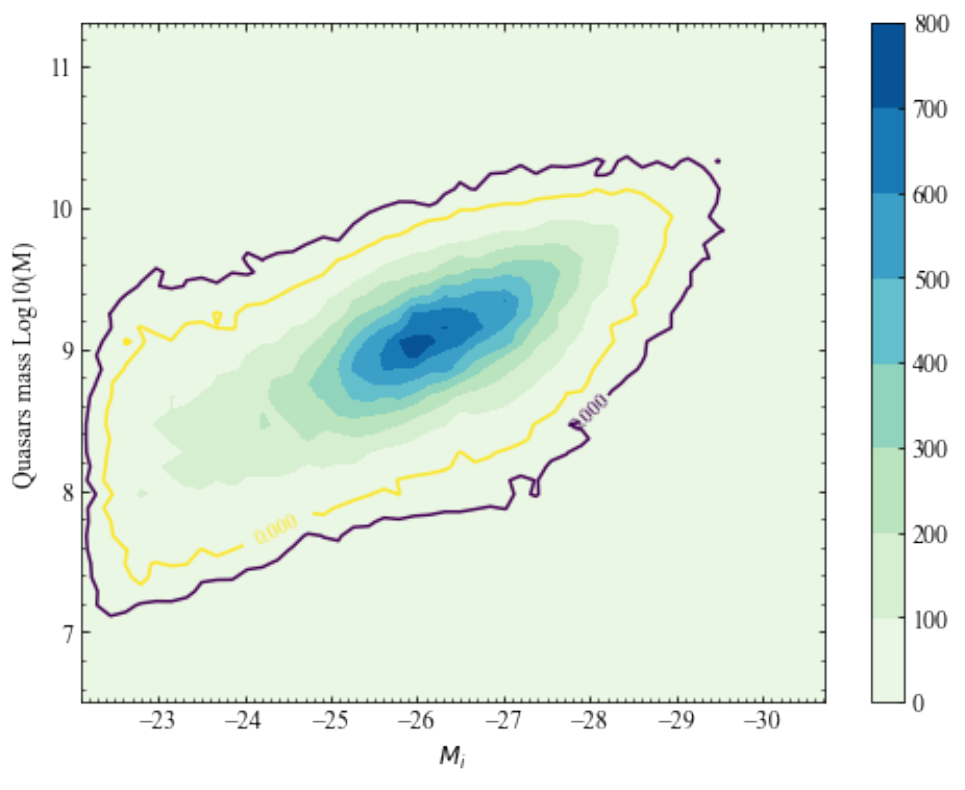}
 \caption{Distribution of our sample (included in the blue rectangle of Fig.~\ref{image_1})
          on the mass-$M_i$ diagram. We use the
          interpolation method to obtain the contour map. $M_{i}=-22.303$ is the
          ultimately selected limiting magnitude for quasars with a mass larger
          than $10^{9}M_\odot$. The purple line contour corresponds to the $99\%$ density
          range and the orange line corresponds to $98\%$. The color bar marks the level
          of the density.}
  \label{image_2}
\end{figure}

\begin{figure}[h]
\centering
 \includegraphics[scale=0.5]{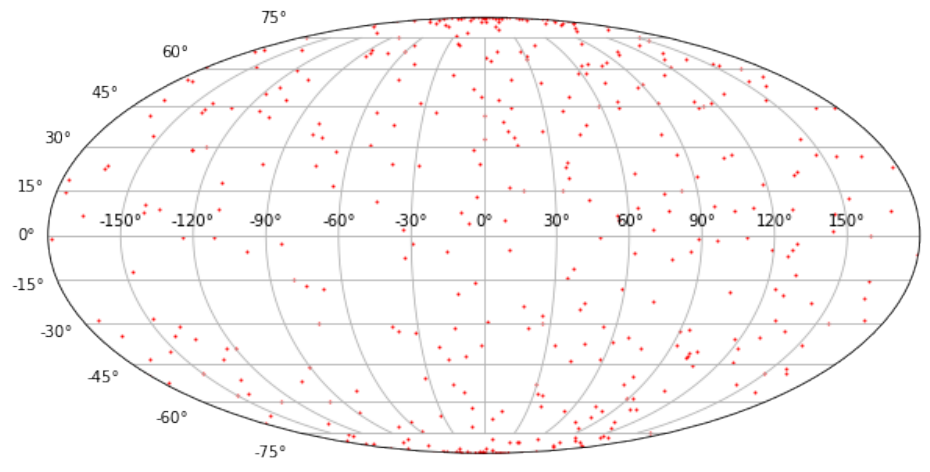}
 \caption{The distribution of our sample (i.e., the objects within the blue rectangle in Fig~\ref{image_1}) on the sky. }
  \label{image_3}
\end{figure}

For the magnetic monopole model, the sample selection is slightly
different from that of the accretion disk model. We first need to
select samples with different mass intervals and then perform
K-correction to make them as complete as possible in a specific
mass interval. Since the sample matrix is larger than 10000
$\times$ 10000, we cannot draw the contour directly. Therefore, we
adopt the interpolation method. We divide the sample into a 70
$\times$ 70 matrix and obtain the number that falls into each grid from
the two-dimensional histogram; then, we can draw the contour map.
The limiting magnitude of the sample we screened is
$M_{i}=-22.303$. Figure \ref{image_2} shows the contour map.
According to the limiting magnitude and the K-correction line, we
finally select a complete sample with mass larger than
$10^{9}M_\odot$ (see the rectangle in Figure \ref{image_1}).
We plot the distribution of these objects on the sky in
Figure \ref{image_3}. Figure \ref{image_3}
shows that the distribution of these objects on the sky is uniform
and isotropic.

\section{Data analysis and results}\label{sec4}

Following the work of Peng et al.
\citep{1986ASSL..121..663P,1998ASPC..151..119P,2001ApJ...551L..23P},
we can model the evolution of massive celestial bodies containing
magnetic monopoles.

In the magnetic monopole quasar model, protons can decay through
catalytic reactions which release a considerable amount of energy.
With the help of magnetic monopoles, mechanical equilibrium can be
maintained at the galactic center, i.e.,
\begin{equation}
   -\frac{dP_{r}}{dr}-\frac{dP_{g}}{dr}=\frac{GM_{r}}{r^{2}}\rho(r),
\end{equation}
where $r$ is the radius, $\rho(r)$ is the density at $r$,
$M_{r}$ is the total mass within $r$, $P_{g}=nkT$ is the gas
pressure, and $P_{r}$ represents radiation pressure. Both $P_{g}$
and $P_{r}$ are enhanced by the release of energy from the
catalytic reaction induced by magnetic monopoles. The radiation
pressure can be calculated as
\begin{equation}
 -\frac{dP_{r}}{dr}=\frac{\kappa \rho(r)L(r)}{4\pi r^{2}c},
\end{equation}
where $\kappa$ is the opacity of matter, $L(r)$ is the
total luminosity within $r$, and $c$ is the speed of light. We assume
that the matter is mainly composed of He$^{4}$ and the opacity is
due to its free free transition. Thus,

 \begin{equation}
     \left \langle \kappa_{ff} \right \rangle_{r}=\kappa_{0}f_{s}\rho(r)T(r)^{-3.5}
     {\rm cm}^{2}/{\rm g},
 \end{equation}
 where $\kappa_{0}=3.77\times 10^{22}$, and $f_{s}\equiv
\sum_{z}\frac{z^{2}g_{ff}x_{z}}{A_{z}} \sim 1$.

The luminosity due to the decay of protons catalyzed by magnetic
monopoles is \citep{1985KexT...30.1056P}
\begin{equation}
    L_{m}(r)\simeq 4\pi m_{B}c^{2}\int\limits_{0}^{r} n_{B}(r)n_{m}(r)c \left \langle \sigma \beta \right \rangle
    r^{2}dr,
\end{equation}
where $n_{m}(r)$ is the number density of monopoles,
$n_{B}(r)$ is the number density of nucleons, and $\beta c$ is the
velocity of nucleons relative to monopoles. $\left \langle \sigma
\beta \right \rangle \simeq 10^{-34} {\rm cm}^{2}$ is the
catalytic reaction rate. Following \citet{1998ASPC..151..119P},
the luminosity of proton decay induced by magnetic monopoles can
be derived as
\begin{equation}
    L=A_{L} \xi x M_{8},
\end{equation}
where $A_{L} = 2.1 \times 10^{44}$ erg/sec, $x = \rho_{c} / (1.84
{\rm g}/{\rm cm}^{3})$, and $M_{8} = M / 10^{8} M_\odot $.  $\xi
\equiv (\zeta /\zeta_{n})(\left \langle \sigma \beta \right
\rangle/10^{-27} {\rm cm}^{2})$ is the ratio of the number of
monopoles to baryons. This parameter varies with the evolution of
the quasar mass. $\zeta$ is the number of monopoles in the
universe, and $\zeta_{n}$ is the Newtonian saturation content of
monopoles in stellar objects. Therefore, the mass loss rate is

\begin{equation}
    \frac{dM}{dt}=L/c^{2}\approx 3\times10^{8}\zeta x_{0}M_{8}^{(1+\alpha)} M_\odot/10^{10}
    Yr.
    \label{equ:4}
\end{equation}

In our study, we take $x_{0}\simeq 1$, $\alpha=1$. For quasars
with different initial masses ($M_{0}$), the parameter $\xi$ is \citep{1997Ap&SS.257..149P}: \\
\begin{equation}
    \zeta \simeq \left\{\begin{matrix}10^{-2}-10^{-3}, M_{0}\approx 10^{12}M_{\odot}
 \\10^{-3}-10^{-4},M_{0}\approx 10^{11}M_{\odot}
 \\10^{-4}-10^{-5},M_{0}\approx 10^{10}M_{\odot}
 \\10^{-5}-10^{-6},M_{0}\approx 10^{9}M_{\odot}
 \\10^{-6}-10^{-7},M_{0}\approx 10^{8}M_{\odot}
\label{equ:11}
\end{matrix}\right.
\end{equation}

In Equation \ref{equ:4}, there is only one variable, $\zeta$. To
calculate the primordial quasar mass, we must determine the value
of $\zeta$ first.  A different initial mass of $M_0$
corresponds to a different $\zeta$. The exact value of $\zeta$
cannot be determined directly from the currently measured mass of
the quasar,  which resides at a redshift of $z$. In other words, $\zeta$
and the initial mass ($M_0$) are coupled with each other.
Therefore, we divide the sample into different mass intervals. Twenty-four
random mass intervals are adopted here, including $10^{7}-10^{8}
M_{\odot}$, $10^{8}-10^{9} M_{\odot}$, $10^{9}-10^{10} M_{\odot}$,
$10^{10}-10^{11} M_{\odot}$, $10^{7}-10^{9} M_{\odot}$, etc. We
use different values of $\zeta$ to try to calculate the initial
mass ($M_{0}$) of the quasar for each mass interval. The derived
initial mass values are then compared with $\zeta$ to see if they are
compatible with each other (cf. Equation \ref{equ:11}). If the initially
assumed $\zeta$ is compatible with the derived $M_{0}$, then the
self-consistency condition is met, and the sample from this mass
interval can be used for further statistics.
Using this method, we finally find that the self-consistent
parameter set should be $\zeta \simeq 10^{-4}-10^{-5}$ and $M >
 10^{9}M_{\odot}$.   

We then draw a horizontal line at the quasar mass of
$10^{9}M_{\odot}$ in Figure \ref{image_2}. It crosses the $99\%$
contour curve at a left point and a right point. The left point
represents the minimum magnitude for the samples with masses larger
than $10^{9}M_{\odot}$, which is $M_{i}$ = -22.303. As shown in
Figure \ref{image_1}, based on the K-correction line of the
limiting magnitude, we can obtain a complete sample. 

\begin{figure}[h]
\centering
 \includegraphics[scale=0.6]{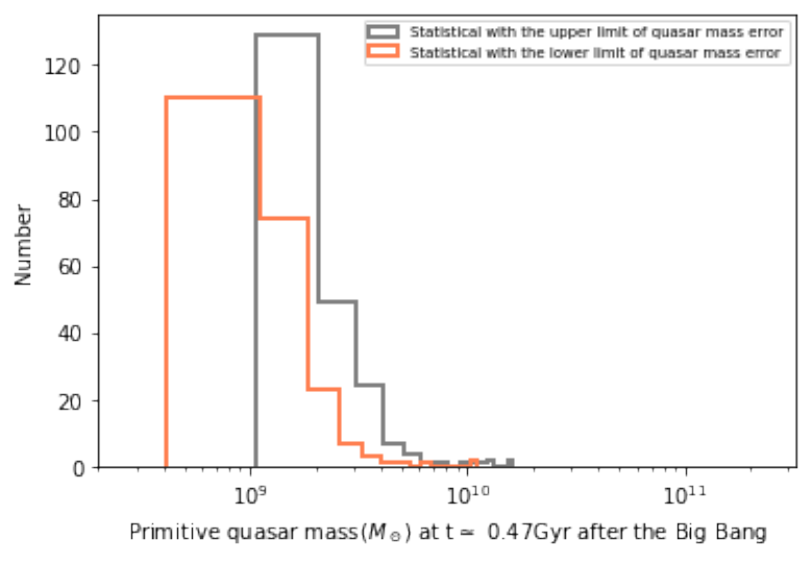}
 \caption{The primitive quasar mass distribution at $t \simeq 0.47$ Gyr after the Big Bang,
          derived from the magnetic monopole model. The grey color histogram shows the result
          of upper limits, and the coral color indicates the result of lower limits. }
  \label{image_4}
\end{figure}

Contrary to the black hole accretion disk model where we must
subtract the accreted mass to obtain the primordial mass
distribution, in the magnetic monopole model, the quasars always
lose mass due to radiation loss. Therefore, we need to add back
the mass loss caused by the magnetic monopole catalysis to obtain
the distribution of the primordial quasar mass. Figure
\ref{image_4} shows our final results. We see now that the
distribution is basically a Gaussian function.


\begin{figure}[h]
\centering
 \includegraphics[scale=0.5]{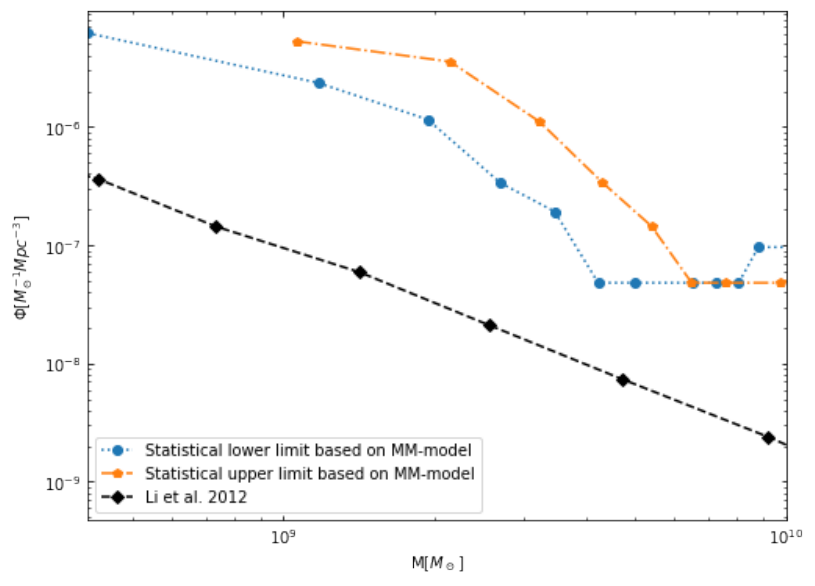}
 \caption{The initial mass function derived from the magnetic monopole model. The abscissa is
 the mass of quasars, and the ordinate is their corresponding numbers. The orange line and the blue line
 correspond to the upper and lower limits of the quasar mass, respectively. The black line represents the initial mass function inferred from observations on
 the local supermassive black holes \citep{2012ApJ...749..187L}.}
  \label{image_5}
\end{figure}

In order to compare with observations, we no longer uniformly
transform the objects to a fixed redshift of 10. On the contrary,
we calculate the mass distribution of the sample 10 billion years
ago.Due to the expansion of the universe, the co-moving
distances of these objects are reduced by nearly 11 times ($D
\propto a(t) \propto \frac{1}{1+z}$). This implies that the selection of our sample is
reasonable.  We then recalculate the initial mass function (IMF)
and compare it with the observed one. The results are
plotted in Figure \ref{image_5}, which shows that the initial masses
are generally higher than the observed values. In the small mass
section of $\sim 10^{8} M_\odot$, this feature is even more
obvious. Due to the limited sensitivities of our telescopes, there
are still many undetected black holes. Our model predicts that
there may be more unobserved supermassive black holes than
previously expected. The new-generation James Webb Space Telescope
may be able to detect more faint quasars and validate our theory.

\section{Conclusion}\label{sec5}

In this study, we use SDSS quasars to analyze the monopole
mass evolution model and obtain the initial mass distribution of
quasars. In this framework, supermassive quasars were formed in
the early universe with a mass of $10^{10}M_\odot$ --
$10^{12}M_\odot$ at a redshift of $z=10$ -- 20 \citep{2014MNRAS.438.1242L}.
According to Peng et al. \citep{1985KexT...30.1056P,1997Ap&SS.257..149P,2001ApJ...551L..23P},
magnetic monopoles may exist in the early universe after the
inflation. These monopoles sunk to the centers of quasars. The
masses of these quasars can then decrease due to continuous
catalytic proton decay induced by these magnetic monopoles
\citep{1988RPPh...51..189R}. This model can explain the existence
of some low mass quasars ($\approx 10^{6} M_\odot$) in the local
universe, such as the central supermassive object of our galaxy.
We use the magnetic monopole quasar model to revisit the initial
mass of quasars at the early universe (for example, 0.47 Gyr after
the Big Bang). We can  obtain a reasonable Gaussian distribution for
the quasar masses. At the same time, we also use the SDSS data to
calculate the initial mass distribution function of quasars based on
the magnetic monopole model and compare it with the observations.
It is found that the theoretical IMF is slightly higher than
previously expected. We predict that there should exist more
supermassive black holes that could be observed in the future.\par

Our theoretical model is also consistent with Soltan's result of
$L\propto M^{2}$ \citep{1980ApJ...238..800A}(see Equation
\ref{equ:4}). In addition, the magnetic monopole model has
successfully predicted positron emissions from the Milky Way
\citep{2001ApJ...551L..23P}. It is predicted that the Milky Way
galaxy should produce and emit many positions, with a production
rate of about $6\times10^{42} e^{+}s^{-1}$. The results are in
good agreement with the observations of (3.4 --- 6.30)
$\times10^{42}e^{+}s^{-1}$ in 2003 \citep{2003A&A...411L.457K}. At
the same time, it is also anticipated that the power of high-energy
photons ($> $0.511 MeV) should be much larger than that of
both the positron annihilation line and the thermal luminosity of
the central compact object. This prediction is also consistent
with observations \citep{1997Ap&SS.257..149P}. Peng et al.
predicted a magnetic field of $B \approx 10-50$ mG at $r = 0.12$
pc from the galaxy center, which is roughly consistent with the
lower limit of 8 mG observed in 2013 \citep{2013Natur.501..391E}.
According to Peng et al., the peak frequency of the
thermal radiation from the supermassive object at our galaxy's
center is approximately $10^{13} Hz$. This is also fairly consistent with
recent observations ($10^{12} Hz$) \citep{2003ApJ...591..891B}.
Finally, as shown in this study, this model provides a reasonable
result on the initial mass distribution. Observational clues
pointing toward the existence of magnetic monopoles are
accumulating. We hope that additional observational evidence
will test the correctness of the theory in the future.


\section{Acknowledgements}
We thank the anonymous referee for useful comments and
suggestions. The authors are thankful for beneficial
discussions with Professor Yong-Feng Huang. We also thank Nanjing
University and Purple Mountain Observatory for valuable support.
\section{Statements $\&$ Declarations}
\subsection{Author Contribution}
Zheng Li is the Principal Investigator and the corresponding author.  Other authors contributed to the interpretation of the results.
\subsection{Funding information}
This work was supported by the Regional Collaborative Innovation Project of Xinjiang Uyghur Autonomous Region (2022E01013), the National Natural Science Foundation of China (12173078 and 11773062), and the West Light Foundation of the Chinese Academy of Sciences (2017-XBQNXZ-A-008).
\subsection{Data availability}
The quasar data used in this article is from \citep{2011ApJS..194...45S}.
\subsection{Conflicts of interest}
The authors declare that they have no conflicts of interest.

\bibliography{reference}{}
\bibliographystyle{apalike}



\end{document}